# Anomalous Mn depth profiles for GaMnAs/GaAs(001) thin films grown by molecular beam epitaxy

J. F. Xu and P. M. Thibado[a]
*Department of Physics, University of Arkansas, Fayetteville, Arkansas 72701*

C. Awo-Affouda, F. Ramos, and V. P. LaBella
*College of Nanoscale Science and Engineering, University at Albany-SUNY, Albany, New York 12203*

Mn concentration depth profiles in Mn-doped GaAs thin films grown at substrate temperatures of 580 and 250 °C using various Mn cell temperatures have been investigated with dynamic secondary ion mass spectrometry and Auger electron spectroscopy. When the samples are grown at a low substrate temperature of 250 °C, the Mn distributes uniformly. For the samples grown at a high substrate temperature of 580 °C, the concentration depth profiles are easily fitted with a temperature-dependent Fermi function only if the Mn concentration is above the solubility limit. However, when the Mn concentration is below the solubility limit, unexpected peaks are observed in the concentration depth profiles.

[a] Electronic mail: thibado@uark.edu

## I. INTRODUCTION

Diluted magnetic semiconductors (DMSs) have been proposed for use in spin-based electronics, or spintronics, because of their potential as both injectors and filters for spin-polarized carriers.[1,2] The novel materials are formed by the substitution of a small fraction of host atoms with magnetic atoms.[3,4] Uniquely, the interaction between magnetic moments is mediated by hole carriers,[5] and the hole carriers are also introduced by the same magnetic species.[6,7]

Recently, more attention has been paid to the III-V magnetic semiconductors such as GaMnAs and InMnAs grown by molecular beam epitaxy (MBE).[1,8] Specifically, the discovery of ferromagnetism at 110 K in GaMnAs has led to intense experimental[9] and theoretical research.[10,11] A Curie temperature higher than room temperature in GaMnAs is theoretically predicted and this would be best for applications. However, experimental progress is severely hindered by the lack of understanding in the doping mechanism and difficulty in growth control[12] even though the quality of GaMnAs films grown by MBE has improved greatly compared to early efforts.[13]

An overall goal has always been to achieve a uniform distribution of the dopant material throughout the thin film.[14] Surprisingly, recent work on Mn-implanted Si shows that Mn concentration fluctuations may be favorable to the ferromagnetic behavior.[15] Regardless of what is best, it is now clear that one must fully understand the Mn doping properties including, diffusion, segregation, and distribution in order to make more progress.

Mn concentration depth profiles in GaMnAs thin films have been studied using Rutherford backscattering spectrometry (RBS).[16] These results also indicate that the distribution is sometimes not uniform. The tricky part with RBS is that it provides a convoluted spectrum that requires modeling to obtain the individual component profiles for Mn, Ga, and As. Secondary ion mass spectrometry (SIMS) can be an excellent technique for determining the Mn concentration depth profiles in samples. Very few studies exist, however. One such study shows SIMS profiles having the expected uniform distribution outcome for GaMnAs films grown at low substrate temperatures.[14] In this article, we present the Mn depth profiles for four cases: above and below the Mn solidsolubility limit and for both high and low substrate temperatures.

## II. EXPERIMENT

Samples were prepared in an ultrahigh vacuum (~2×10$^{-10}$ Torr) MBE growth chamber (Riber 32) which includes Ga and Mn effusion cells together with a two-zone As valved-cracker cell and a reflection high-energy electron diffraction (RHEED) system operating at 15 keV. Commercially available, "epiready," semi-insulating 2 in. diameter GaAs(001)±0.1° wafers were cleaved into quarters and indium mounted on a 2 in. diameter standard MBE molybdenum block. The substrate was heated to 580 °C while exposing the surface to 10 μTorr As$_4$ to remove the surface oxide layer. A thin buffer layer of GaAs was grown on the substrate for 5 min. During this time RHEED oscillations were used to determine that the growth rate of the GaAs was 780 nm/h. (However, due to our shutter transient the actual growth rate is closer to 650 nm/h.[17,18]) Next, the substrate temperature was set to the desired growth temperature of either 580 or 250 °C and GaMnAs films were grown for 1 h. The sample was then cooled at 1.5 °C/s to 200 °C, removed from the UHV system, and cleaved into multiple smaller pieces (10×10 mm$^2$) for characterization measurements.

The Mn concentration profiles were determined using a dynamic SIMS (Perkin-Elmer PHI 6300 Quadrupole) and Auger depth profiling systems. The sample was sputtered with an argon ion gun using a focused spot size of about 1 mm in diameter. The depths of the sputtered craters were determined using a profilometer (Tencor Alphastep 200). It was assumed that the average sputtering rate of Ga$_x$Mn$_{1-x}$As did not change with Mn concentration. However, in order to accurately calibrate the Mn concentrations by SIMS, a standard sample was prepared. For this a semi-insulating GaAs substrate was implanted with Mn ions extracted from a plasma arc source at an energy of 300 keV using a dose of 5×10$^{15}$ cm$^{-2}$.

## III. RESULTS

The SIMS Mn concentration depth profiles for the GaMnAs thin films grown using the low substrate temperature (250 °C) and various Mn cell temperatures are shown in Fig. 1. The Mn concentration in each film is fairly uniform, and the concentration increases with increasing Mn cell temperature, as expected. The samples have a Mn concentration profile that is shaped like a step. The concentration drops at the depth of around 620 nm for samples grown at Mn cell temperatures of 700, 750, and 800 °C. However, the step depth increases to about 800 nm for the sample grown using a Mn cell temperature of 900 °C. At this Mn cell temperature the Mn concentration is now high enough to alter the film thickness. Note, for the Mn cell temperature of 900 °C the Mn concentration is too high to obtain a proper concentration calibration using SIMS, so Auger experiments were performed to calibrate the SIMS concentration throughout the SIMS profile.

The SIMS Mn concentration depth profiles in the GaMnAs thin films grown using the high substrate temperature (580 °C) and various Mn cell temperatures are shown in Fig. 2. The depth profiles are clearly different from those shown in Fig. 1. For the samples grown at higher Mn cell temperatures (800, 850, and 900 °C), the Mn concentration is higher near the surface and then drops gradually with depth. However, for the samples grown at Mn cell temperatures of 700 and 750 °C, a Mn concentration peak appears around 670 nm.

A comparison of Mn depth profiles using a high Mn cell temperature of 900 °C but different substrate temperatures is shown in Fig. 3. When the thin film was grown at 580 °C the Mn depth profile is shown in Fig. 3(a). Here the Mn concentration starts at about 7×10$^{21}$ cm$^{-3}$ then monotonically decreases to zero at a depth of approximately 1200 nm. The depth at which the Mn concentration drops to half of its starting value is 355 nm. When the thin film was grown using the lower substrate temperature (250 °C), the results are shown in Fig. 3(b). This profile does begin with a significant peak at the surface, but after this it stays nearly flat at about 2×10$^{21}$ cm$^{-3}$. The Mn concentration drops to half of its starting value at a depth of 855 nm. Next, the Mn level abruptly drops two orders of magnitude and decreases to zero at around 1000 nm.

Due to the smooth and monotonic behavior of both the SIMS profiles in Fig. 3, we were able to fit them with Fermi functions which are shown as thin lines in Fig. 3. This phenomenological fitting function was chosen primarily because it provides an excellent fit to the data. However, this function also captures the exponential relationship that atomic diffusion has with temperature. The functional form we are using is shown as follows:

$$[Mn(x)]/[Mn(0)] = 1/\{\exp[F(x-x_F)/k_BT] + 1\}$$
,
where $[Mn(x)]$ is the Mn concentration as a function of depth $x$, $[Mn(0)]$ is the starting Mn concentration, $k_B$ is Boltzmann's constant, $T$ is absolute substrate temperature during growth, $x_F$ is depth at which the Mn concentration falls to half of its original value, and $F$ is the only fitting parameter. A good fit to the data of Fig. 3(a) is obtained using $F$=0.5 meV/nm, while a good fit to the data of Fig. 3(b) is obtained using $F$=1.5 meV/nm.

A comparison of Mn depth profiles grown using a low Mn cell temperature of 700 °C, but different substrate temperatures is shown in Fig. 4. At a high substrate temperature (580 °C) the SIMS profiles have a sharp peak near 670 nm, as shown in Figs. 4(a) and 4(b). In addition to the sharp peak, there is a surprising drop in the concentration by about a factor of 10 between 450 and 650 nm. When the thin film is grown using a low substrate temperature

(250 °C), yet still having the Mn cell temperature at 700 °C, the results are much different as shown in Fig. 4(c). Here, roughly speaking, the Mn concentration is nearly uniform over the entire range. It is interesting to notice that the Mn concentration is lower in this sample when compared to the one grown at a higher substrate temperature (580 °C). However, there is a corresponding broader profile for the Mn distribution in the low substrate temperature (250 °C) film.

## IV. DISCUSSION

In this section, we explain our results starting with Fig. 1. Notice that the 900 °C profile shown in Fig. 1 is about 50% thicker than the rest. Based on the RHEED oscillations and after correcting for the measured shutter transient using the flux gauge, the GaAs growth rate, taken by itself, gives a sample thickness of about 650 nm. The extra thickness is due to codepositing high concentrations of Mn during the GaAs growth. Since the Mn deposition rate is so high, the overall thickness of the sample is larger. SIMS data contain no information about the structural properties or crystalline phase of the Mn, only its concentration as a function of depth. For the lower Mn concentration samples, the Mn has a negligible effect on the overall growth rate and film thickness. More details about the data shown in Fig. 1 will be discussed later when Figs. 3 and 4 are discussed.

Continuing to Fig. 2, notice that the low Mn concentration runs have a spike in the data, while the high Mn concentration runs have a monotonic behavior. The monotonic behavior is addressed first. The primary difference between these data and the data acquired at a substrate temperature of 250 °C is the steep slope of the concentration curves. Even though the Mn atoms are deposited uniformly throughout the growing film, the concentration profile grows continuously from the substrate to the film surface. These data make it clear that the Mn being deposited into the growing film is not being incorporated into the film at the time it is deposited. It is staying on the growth front longer and grading the concentration profile. Thus, one can conclude that the incorporation rate is lower at higher substrate temperatures. Wood and Joyce[19] investigated Sn doping in GaAs films grown by MBE. They found that the Sn incorporation rate decreases when the substrate temperature is increased. This is consistent with these findings as well. More details about the data shown in Fig. 2 will be discussed later when Figs. 3 and 4 are discussed.

A quantitative analysis of the Mn concentration profiles in Fig. 3 reveals a deeper understanding of this system. By just changing the substrate temperature the Mn profile goes from a step function at 250 °C to a more gradual transition at 580 °C. The first attempt to model this change in the data was to only change $T$ in the Fermi function. However, a good fit requires raising the substrate temperature to 2300 °C instead of 580 °C. Obviously, something else must be factored into the analysis. The other factor used was the fitting constant $F$. Meaning can be assigned to the fitting constant after calculating the product $F_xF$. This product is found to be 180 meV for the sample grown at 580 °C and 1300 meV for the sample grown at 250 °C. Since this product is changing by a factor of 7, a new diffusion channel is available for the Mn to move around the sample. An energy of about 1 eV would normally correspond to bulk diffusion, while an energy of about 0.1 eV would correspond to surface diffusion.[20] Therefore, for the higher substrate temperatures the Mn atoms can float longer along the growth front,[21] while at the lower substrate temperature they are less mobile and likely buried more uniformly in the growing crystal.

There is a significant rise in the Mn concentration at the sample surface for the data shown in Fig. 3(b). The interesting thing about this increase in Mn is that it is concentrated only at the surface of the film. It is not a gradual monotonic increase in the Mn concentration as observed with the higher substrate temperature data. The cause of this is likely due to a fixed concentration of Mn continuously floating along the growing surface of the GaMnAs film.

The sharp peak in the low Mn concentration data shown in Figs. 4(a) and 4(b) is more surprising. This peak does correspond to the location inside the sample where the oxide was removed. We believe that the Mn moves to the interface after being deposited in the growing film via bulk diffusion. It is also noteworthy that the Mn peak at the interface has a height of ∼$10^{-19}$ cm$^{-3}$, which is also thought to be the solid solubility limit for Mn in GaAs.[3] Thus, the Mn diffuses to the interface region until the local concentration reaches the solubility limit, at which time the Mn stops flowing to the interface region. The reason why the Mn flows to the interface is unclear, but it is natural to assume that there are defects remaining at this interface, such as oxygen, from the oxide removal process that attracts the Mn.

The Mn concentration is nearly zero adjacent to the interface in Figs. 4(a) and 4(b), which is also consistent with the bulk diffusion idea just discussed. Such a region is often called a denuded zone. Denuded zones form when the atoms are deposited in a region next to an interface and then they diffuse to the interface and get captured by it. The width of the denuded zone is related to the rate of diffusion. In order to quantify the diffusion rate, one would need to measure the denuded width for a few substrate temperatures,[22] which was not possible for this set of experiments.

A small peak at the interface is observed when the substrate temperature is lowered to 250 °C, as shown in Fig. 4(c). The lower substrate temperature does lead to a lower Mn mobility and therefore we believe this is the reason why there is no denuded zone. As before with the high substrate temperature data, this small interface peak has a height equal to the solubility limit of $\sim 10^{-19}$ cm$^{-3}$ for Mn in GaAs. Even though the low temperature profile is very different from the high temperature profile, the total amount of Mn deposited is similar. A broader but lower concentration for the low temperature sample. For the high temperature sample, the Mn has redistributed into the two sections with the denuded zone in between. Notice that the Mn concentration has a slight rise from the surface toward the bulk for all three of the runs shown in Fig. 4. We believe that this is a SIMS artifact related to preferential sputtering. The idea is that some of the Mn hit by the sputtering beam is actually knocked into the film and results in the slight increase in the concentration. This effect is small and would not be noticed in the high concentration runs.

When one compares the different Mn concentrations for the high substrate temperature growths, we do not see the tall interface peak for the higher concentration samples (see Fig. 2, Mn cell temperatures of 800, 850, and 900 °C). We believe that this is because the Mn concentration is so high that it immediately exceeds the solubility limit. This leads to the simpler monotonic depth profile behavior.

## V. CONCLUSIONS

In conclusion, Mn-doped GaAs thin films have been successfully grown by MBE at both high (580 °C) and low (250 °C) substrate temperatures. Furthermore, the role of doping Mn above and below the solubility limit on the depth profiles has been studied. For the samples grown at a high substrate temperature but below the solubility limit, a Mn concentration peak near the film/substrate interface is observed. Whereas, when the Mn concentration is higher than the solubility limit the Mn diffuses to the surface (not the interface) by floating along the growth front. Overall, this systematic, broad Mn concentration, and substrate temperature range study shows numerous ways the Mn concentration profile is altered. These properties need to be considered when designing DMS structures.

## ACKNOWLEDGMENTS

The authors gratefully acknowledge Richard Moore for the Auger electron spectroscopy analysis. The authors would like to acknowledge the support for this work from National Science Foundation under Grant Nos. DMR-0405036 and DMR-Career-0349108, and Marco Interconnect Focus Center.

[1] H. Ohno, Science **281**, 951 (1998).

[2] S. J. Potashnik, K. C. Ku, R. Mahendiran, S. H. Chun, R. F. Wang, N. Samarth, and P. Schiffer, Phys. Rev. B **66**, 012408 (2002).

[3] R. Moriya and H. Munekata, J. Appl. Phys. **93**, 4603 (2003).

[4] K. M. Yu, W. Walukiewicz, T. Wojtowicz, I. Kuryliszyn, X. Liu, Y. Sasaki, and J. K. Furdyna, Phys. Rev. B **65**, 201303 (2003).

[5] T. Dietl, H. Ohno, and F. Matsukura, Phys. Rev. B **63**, 195205 (2001).

[6] M. Bolduc, C. Awo-Affouda, A. Stollenwerk, M. B. Huang, F. G. Ramos, G. Agnello, and V. P. LaBella, Phys. Rev. B **71**, 033302 (2005).

[7] K. M. Yu, W. Walukiewicz, T. Wojtowicz, W. L. Lim, X. Liu, U. Bindley, M. Dobroolska, and J. K. Furdyna, Phys. Rev. B **68**, 041308 (2003).

[8] H. Ohno, A. Shen, F. Matsukura, A. Oiwa, A. Endo, S. Katsumoto, and Y. Iye, Appl. Phys. Lett. **69**, 363 (1996).

[9] R. P. Campion, K. W. Edmonds, L. X. Zhao, K. Y. Wang, C. T. Foxon, B. L. Gallagher, and C. R. Staddon, J. Cryst. Growth **247**, 42 (2003).

[10] T. Dietl, H. Ohno, F. Matsukura, J. Cibert, and D. Ferrand, Science **287**, 1019 (2000).

[11] T. Jungwirth *et al.*, Phys. Rev. B **72**, 165204 (2005).

[12] M. Malfait, J. Vanacken, V. V. Moshchalkov, W. Van Roy, and G. Borghs, Appl. Phys. Lett. **86**, 132501 (2005).

[13] M. Llegems, R. Dingle, and L. W. Rupp, J. Appl. Phys. **46**, 3059 (1975).

[14] L. X. Zhao *et al.*, Semicond. Sci. Technol. **20**, 369 (2005).

[15] M. Bolduc, C. Awo-Affouda, F. Ramos, and V. P. LaBella, J. Vac. Sci. Technol. A **24**, 1648 (2006).

[16] J. L. Hilton, B. D. Schultz, S. Mckernan, and C. J. Palmstrom, Appl. Phys. Lett. **84**, 3145 (2004).

[17] T. E. Harvey, K. A. Bertness, R. K. Hickernell, C. M. Wang, and J. D. Splett, J. Cryst. Growth **251**, 73 (2003).

[18] Ch. Heyn and S. Cunis, J. Vac. Sci. Technol. B **23**, 2014 (2005).

[19] C. E. C. Wood and B. A. Joyce, J. Appl. Phys. **49**, 4854 (1978).

[20] J. L. Hilton, B. D. Schultz, S. McKernan, S. M. Spanton, M. M. R. Evans, and C. J. Palmstrom, J. Vac. Sci. Technol. B **23**, 1752 (2005).

[21] Q. Gong, R. Notzel, J. H. Wolter, H.-P. Schonherr, and K. H. Ploog, J. Cryst. Growth **242**, 104 (2002).

[22] H. Yang, V. P. LaBella, D. W. Bullock, Z. Ding, J. B. Smathers, and P. M. Thibado, J. Cryst. Growth **201/202**, 88 (1999).

FIG. 1. Mn concentration SIMS depth profiles for Mn-doped GaAs thin films grown by MBE using a low substrate temperature, $Ts$=250 °C, and four different Mn cell temperatures: 700, 750, 800, and 900 °C.

FIG. 2. Mn concentration SIMS depth profiles for Mn-doped GaAs thin films grown by MBE using a high substrate temperature, $Ts$=580 °C, and five different Mn cell temperatures: 700, 750, 800, 850, and 900 °C.

FIG. 3. Mn concentration SIMS depth profiles for Mn-doped GaAs thin films grown by MBE using two different substrate temperatures, $Ts$, but each was grown using the same high Mn cell temperature of 900 °C: (a) $Ts$=580 °C and (b) $Ts$=250 °C. Also shown are fits as thin lines using Fermi functions with fitting parameter, $F$: (a) $F$=0.5 meV/nm and (b) $F$=1.5 meV/nm.

FIG. 4. Mn concentration SIMS depth profiles for Mn-doped GaAs thin films grown by MBE using two different substrate temperatures, $Ts$, but each was grown using a low Mn cell temperature of 700 °C: (a) $Ts$=580 °C (site 1), (b) $Ts$=580 °C (site 2), and (c) $Ts$=250 °C.

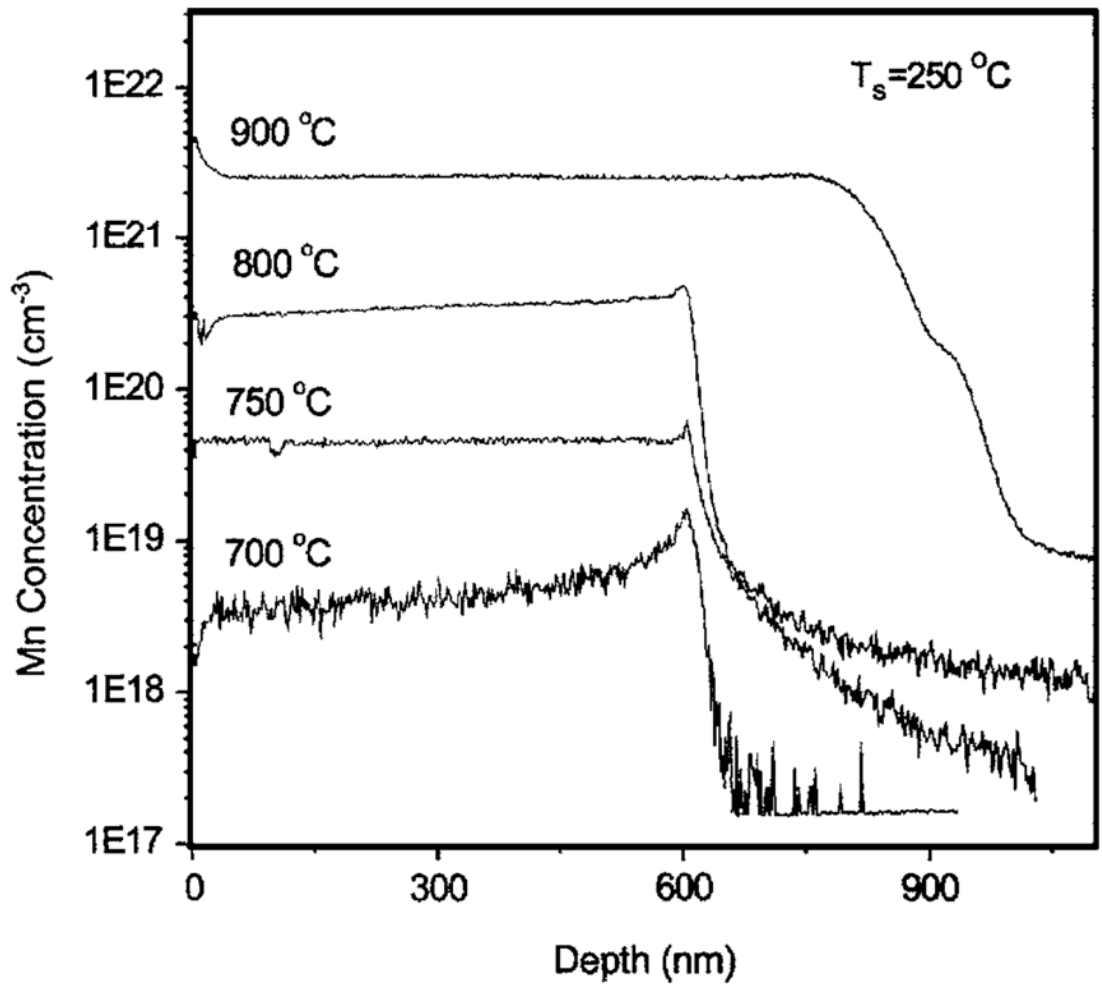

Figure 1.

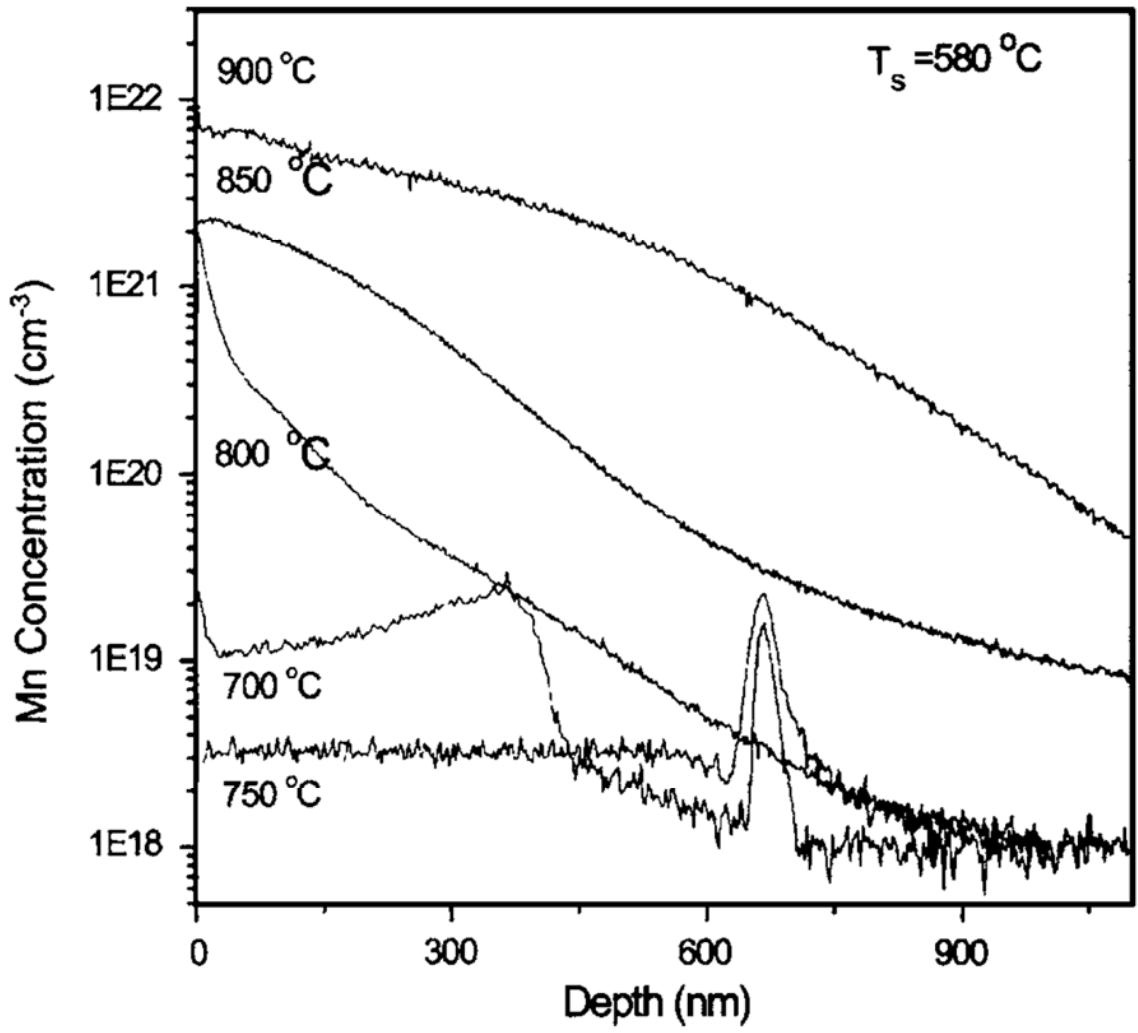

Figure 2.

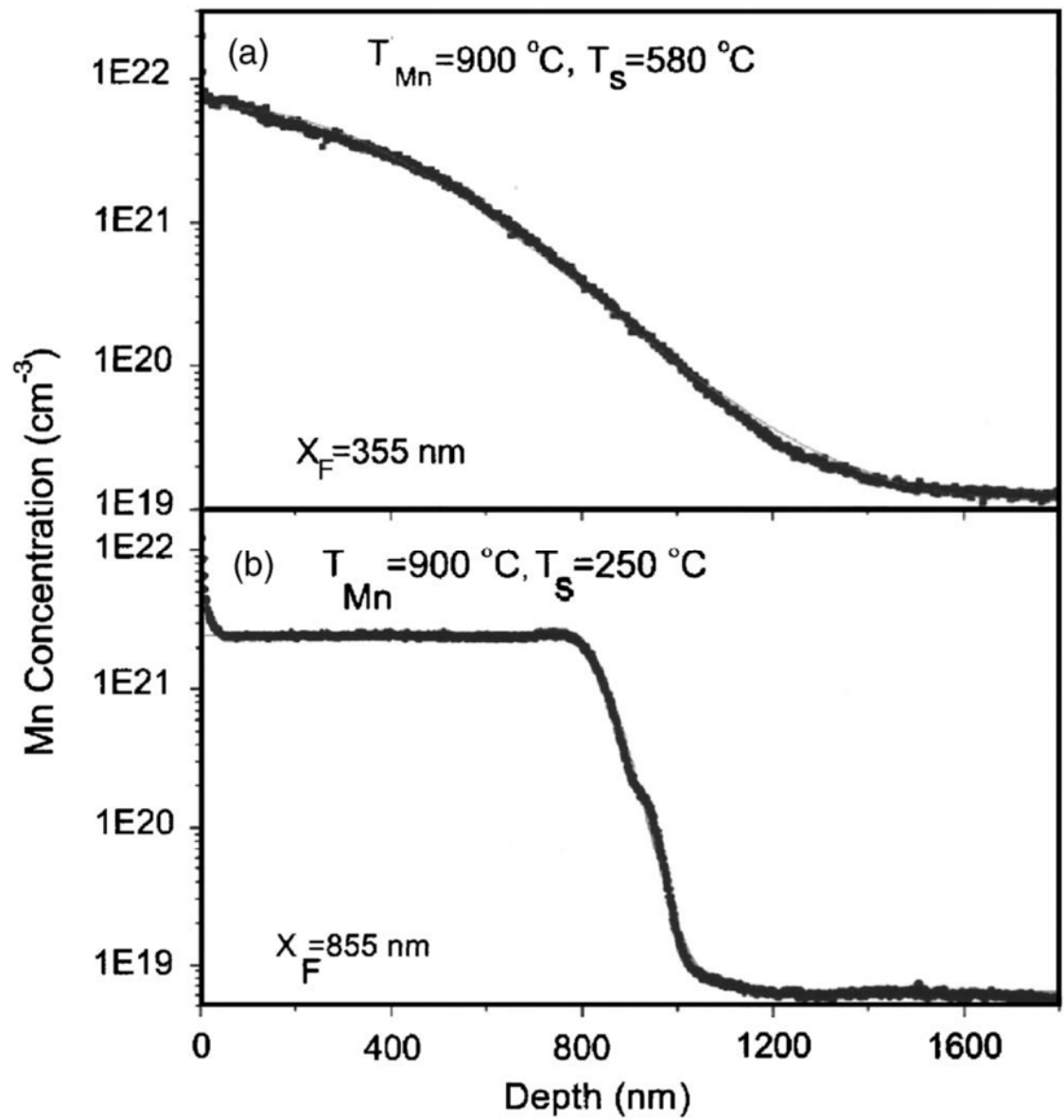

Figure 3.

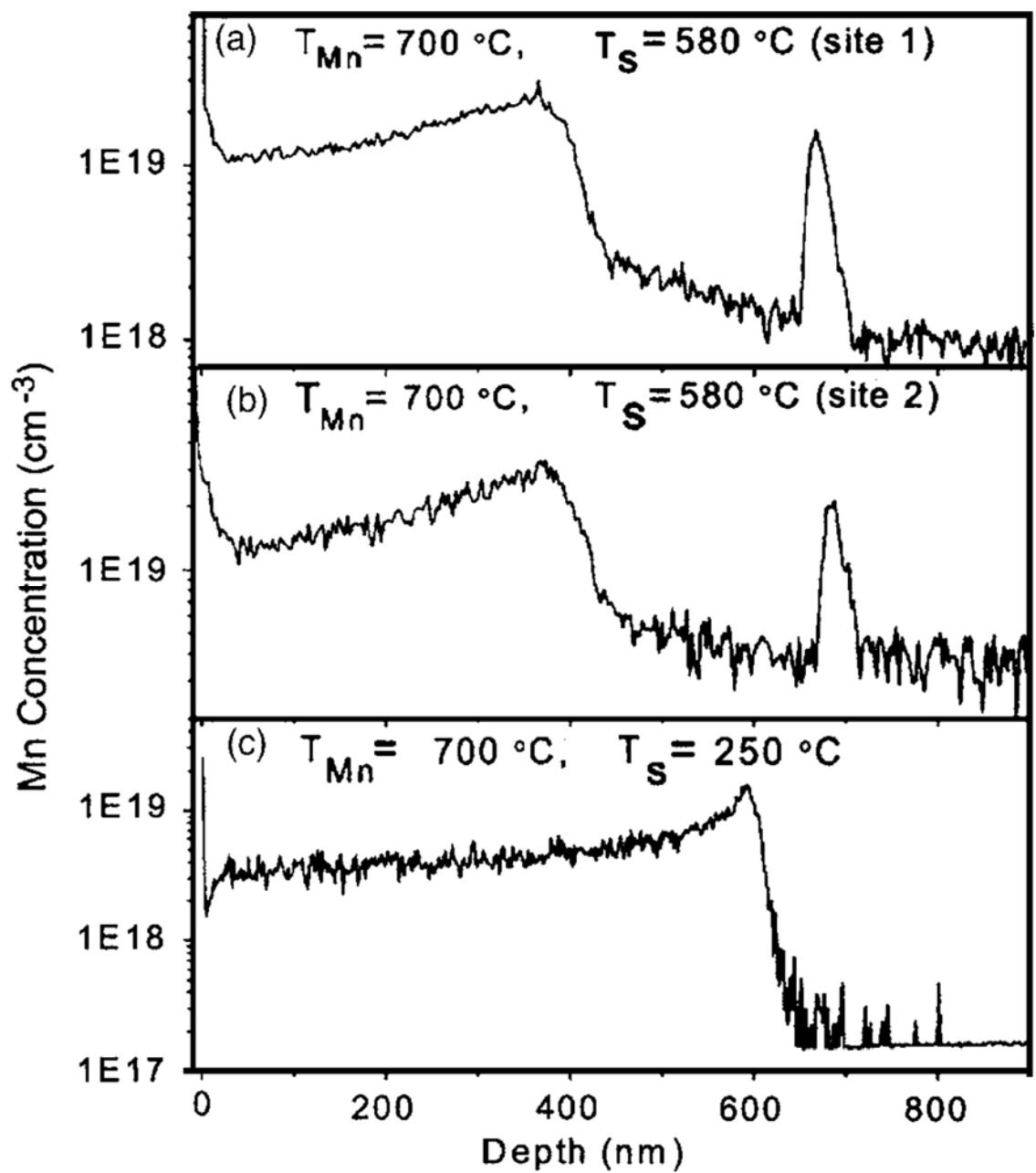

Figure 4.